\documentclass[10pt,aps,prd, twocolumn,superscriptaddress,showpacs,nofootinbib,preprintnumbers]{revtex4-1}
\usepackage[english]{babel}
\usepackage[latin1]{inputenc}
\usepackage{hyperref}
\usepackage{graphicx,color}
\usepackage{latexsym}
\usepackage{amsmath}
\usepackage{amsfonts,dsfont}
\usepackage{amssymb,slashed}
\usepackage{bbold}
\usepackage{verbatim,epigraph}

\DeclareMathOperator{\Tr}{Tr}

\begin{document}

\title{Anomalies and Renormalization of Impure States in Quantum Theories}

\author{\textbf{Kumar S. Gupta}}
\thanks{On leave from SINP, India}
\email[]{kumars.gupta@saha.ac.in}
\affiliation{{Theory Division, Saha Institute of Nuclear Physics,}\\ \emph{1/AF Bidhannagar, Kolkata 700 064, India}}
\affiliation{International Institute of Physics (IIP-UFRN) \\
Av. Odilon Gomes de Lima 1722, 59078-400 Natal, Brazil}

\author{\textbf{Amilcar Queiroz}}  
\email[]{amilcarq@unb.br}

\affiliation{Instituto de Fisica, Universidade de
Brasilia, Caixa Postal 04455, 70919-970, Brasilia, DF, Brazil}


\begin{abstract}
In a Hamiltonian approach to anomalies parity and time reversal symmetries can be restored by introducing suitable impure (or mixed) states. However, the expectation values of observables such as the Hamiltonian diverges in such impure states. Here we show that such divergent expectation values can be treated within a renormalization group framework, leading to a set of $\beta$-functions in the moduli space of the operators representing the observables. This leads to well defined expectation values of the Hamiltonian in a phase where the impure state restores the $P$ and $T$ symmetry. We also show that this RG procedure leads to a mass gap in the spectrum. Such a framework may be relevant for long wavelength descriptions of condensed matter systems such as the quantum spin Hall effect. 
\end{abstract}

\maketitle


Anomalies occur when a classical symmetry is broken due to quantization \cite{Harvey2007}. In quantum field theories, they were first discovered within the perturbative approach \cite{jackiw1969,Adler1969a,Fujikawa1984d}. It was soon realized that anomalies are intrinsically non-perturbative in nature \cite{Adler1969,Alvarez-Gaume1984}, characterized by geometric and topological structures of the theory \cite{AlvarezGaume1984269}. This point of view has led to a much wider realization of anomalies in diverse areas including particle physics \cite{Witten1982324}, black holes \cite{Bimonte1993,Balachandran1996a} and important condensed matter systems such as quantum Hall systems and the associated edge states \cite{CallanJr1985427,Stone198789,Balachandran1992,Balachandran1995b,Balachandran1994a}.

One way to understand anomalies involves the study of unbounded operators \cite{reed1975methods} in quantum theories such as the momentum and the Hamiltonian. The complete definition of such unbounded operators requires the specification of appropriate boundary conditions or equivalently their domains so that they become self-adjoint operators. A symmetry can be implemented in the quantum theory if the corresponding generators leave the domain of the Hamiltonian invariant and then if they commute with the Hamiltonian. Sometimes a symmetry generator does not preserve the domain of the Hamiltonian. In this case, the symmetry is broken due to quantization being thus anomalous. This approach to anomalies \cite{PhysRevD.34.674,jackiw,PhysRevD.66.125013} is very general and it allows its study in a large class of quantum systems such as molecular physics \cite{PhysRevLett.87.220402,ksgpolar}, condensed matter systems \cite{ezawa}, integrable models \cite{biru1,biru2,biru3} and black hole physics \cite{PhysRevLett.87.220402,sen1,sen2}. This approach can also be adapted to quantum field theories \cite{Balachandran2012}, where the mode expansion encodes the information about the appropriate boundary conditions.

In recent papers \cite{Balachandran2012,Balachandran:2011gj}, it has been shown that while a system might admit an anomaly when in a pure state, it is possible to restore the anomalous symmetry when the system is in a suitable impure (or mixed) state. This naturally leads to the question as to what are the expectation values of the observables of the system in such impure state. 

The aim of this Letter is to provide an answer to this question. In particular, we focus our attention to the case of the expectation values of the Hamiltonian. If the impure state is obtained by an appropriate summation of pure states each corresponding to different domains, it is well known that the expectation values of observables such as the Hamiltonian diverge in such an impure state. In this Letter we show that such divergences can be handled by using renormalization group (RG) techniques \cite{jackiw,rajeev}. The corresponding beta function provides a flow in the moduli space of the operators and the choice of the associated impure state depends on the energy scale of the problem. Therefore, the RG flow removes the divergence and even more interestingly it makes the quantum theory scale dependent. Physically it says that if we demand anomalous symmetry restoration, then the appropriate boundary conditions depend on the energy scale at which the system is being probed. To our knowledge this is the first attempt that divergences arising from such impure states are analysed using non-perturbative RG techniques.

We will present these ideas below in the problem of a particle on a circle and its anomalous parity $P$ and time-reversal $T$ symmetries. We first recall how impure states with a non-trivial von Neumann or entanglement entropy can be used to restore such anomalous symmetries. Then we discuss how to obtain expectation values of the observables such as the Hamiltonian in the impure state making explicit the appearance of divergences. We then show how the application of RG techniques lead to finite answer for the expectation values of observables in the impure state. We next provide general arguments to the emergence of such impure states restoring anomalous symmetries. Finally we discuss how this approach can be applied to system with gapped bulk and edge states that are modelled by quantum states of a particle on a circle.

The Hamiltonian for a particle on a circle of radius $R$ is given by
 \begin{equation}
	 \label{hamiltonian-circle}
       H=-\frac{1}{R} \frac{\partial^2}{\partial \varphi^2}.
 \end{equation}
The Hamiltonian $H$ is self-adjoint in the family of domains $D_\theta=\left\{ L^2(S^1,d\varphi):~\psi(2\pi)=e^{i\theta} \psi(0), ~ \psi'(2\pi)=e^{i\theta} \psi'(0)\right\}$, with $\theta\in [0,2\pi)$. The parameter $\theta$ may be associated to the magnetic flux through the circle. In simpler terms, this parameter labels the allowed family of boundary conditions for which the Hamiltonian $H$ is self-adjoint. The dimension of the Hamiltonian is $[\rm{length}]^{-1}$.

For a fixed domain $D_\theta$, the eigenvalues and eigenfunctions of $H$ are given by
\begin{align}
	\label{energy-eigenvalues}
    \epsilon_{\theta,n} = \frac{1}{R}(n+\frac{\theta}{2\pi})^2\equiv \frac{n_\theta^2}{R}, \qquad  \psi_{\theta,n} = \frac{e^{i n_\theta \varphi}}{\sqrt{2\pi}},
\end{align}
with $n\in\mathbb{Z}$.

The action of parity $P$ on the wave-function is given by $[P\psi_\theta](\varphi)=\psi_{-\theta}(\varphi)$. Indeed, since $[P\psi](\varphi)=\psi(2\pi-\varphi)$, we see that $[P\psi_\theta](2\pi)=\psi_\theta(0)=e^{-i\theta}\psi_\theta(2\pi)=e^{-i\theta}[P\psi_{\theta}](0)$. The parity operator therefore changes the boundary condition of the Hamiltonian. Thus, if $\theta\neq 0,\pi$, parity is broken due to quantization being thus anomalous \cite{Balachandran2012}. Also, since time-reversal $T$ is a anti-unitary transformation, then for $\theta\neq0,\pi$ it also changes the domain of the Hamiltonian, $[P\psi_\theta](\varphi)=\psi_{-\theta}(\varphi)$, being therefore anomalous. 

The eigenstates of $(H,D_\theta)$ are pure states. They are mapped to the domain $D_{-\theta}$ by parity $P$ and time reversal $T$. In \cite{Balachandran2012}, it was proposed that if the system is in a phase described by the impure state of the form
\begin{equation}
	\label{parity-impure-state-1}
      \rho =\frac{1}{2}\left( \rho_\theta+ \rho_{-\theta}\right), \quad \textrm{with} \quad \rho_{\theta} = |\psi_\theta\rangle\langle\psi_\theta|,      
\end{equation}
then parity $P$ and time-reversal $T$ are no longer anomalous. Indeed, parity $P$ and time reversal $T$ leave the impure state $\rho$ invariant. Indeed, under these operations $\theta \mapsto - \theta$, but since the impure state $\rho$ is obtained by averaging over these values of $\theta$, it is invariant under $P$ or $T$.

Eigenstates of the self-adjoint operator $H$ form a basis in the Hilbert space. Therefore we may expand any state in an arbitrary domain in terms of the eigenstates in the original domain of the Hamiltonian. Let $\psi_{\theta,n}$ be an eigenstate of $H$ belonging to the domain $D_\theta$. Then $\chi_{\theta,n}=P \psi_{\theta,n}\in D_{-\theta}$, with $\theta\neq 0,\pi$. We may then expand $\chi_{\theta,n}$ in terms of the eigenstates $\{\psi_{\theta,k}\}\subset D_\theta$ as  
      \begin{equation}
            \chi_{n,\theta}=\sum_k~b_{nk}~\psi_{\theta,k},
      \end{equation}
where
\begin{equation*}
      \label{projected-state-1}
      b_{nk} =\int_0^{2\pi} \frac{d\varphi}{2\pi}~e^{-i\left(k+\frac{\theta}{2\pi}\right)\varphi}~e^{i\left(n-\frac{\theta}{2\pi}\right)\varphi} = \frac{1}{2\pi}\frac{e^{-2i\theta}-1}{n-k-\frac{\theta}{\pi}}. 
\end{equation*}
 
 
The expectation value of $H\equiv \left(H,D_\theta\right)$ in the impure state (\ref{parity-impure-state-1}) associated to the $n$-th energy level, that is, $\rho_n=(1/2)(\rho_{\theta,n}+\rho_{-\theta,n})$, is 
 \begin{align}
	\label{exp-val-Hamilton-1}
    \langle H \rangle_n=\Tr~\rho_n H &= \frac{1}{2}\left( \epsilon_{\theta,n} + \sum_k |b_{nk}|^2~\epsilon_{\theta,k} \right), \\
    \textrm{where}\quad |b_{nk}|^2 &= \frac{1}{2\pi^2}\frac{1-\cos 2\theta}{\left(n-k-\frac{\theta}{\pi}\right)^2}.  \label{square-proj}
\end{align}
The second term in the RHS of (\ref{exp-val-Hamilton-1}) diverges in the UV region, that is, when $k \rightarrow \infty$. To see this, we first convert $\sum_k~\to~\int dk$, so that 
\begin{equation}
  \sum_k |b_{nk}|^2~\epsilon_k ~ \to ~ I(n,\theta)=\int_{-\infty}^\infty dk ~ |b_{nk}|^2~\epsilon_k.
\end{equation}
We then use (\ref{square-proj}) together with (\ref{energy-eigenvalues}), to get
\begin{align}
	\label{I-integral}
  I(n,\theta)&=\frac{1-\cos 2 \theta}{2\pi^2~R}\int_{-K}^K dk ~ \frac{k_\theta^2}{\left(n_{-\theta}-k_\theta\right)^2} \\
  &=2K-\frac{2 K n_{-\theta}^2}{K^2-n_{-\theta}^2}+2n_{-\theta}\log\left|\frac{K-n_{-\theta}}{K+n_{-\theta}} \right|, \nonumber
\end{align}
where we have introduced a UV cut-off $K$. We observe that the first term in the RHS is linearly divergent while the remaining terms vanish as $K\to\infty$. Therefore (\ref{I-integral}) becomes effectively 
\begin{equation}
  I(n,\theta)=\frac{1}{\pi^2}\lim_{K\to \infty} \frac{1-\cos 2\theta}{R}~K.
\end{equation} 
The expectation value of the Hamiltonian is thus
\begin{align}
	\label{n-level-energy}
 \langle H \rangle_n &= \frac{1}{2R} \left[  n_{\theta}^2 + \lim_{K\to\infty}\frac{1-\cos 2 \theta}{\pi^2}~ K \right].  
\end{align}

Observe that the second term on the RHS of (\ref{n-level-energy}), which is divergent, does not depend on the level $n$. Indeed, the UV divergence occurs solely due to the projection of a state in the domain $D_{-\theta}$ in terms of the domain $D_\theta$. Thus even though the impure state removes the $P$ and $T$ anomaly, expectation values of the observables such as the energy calculated in the impure state appear to diverge. This is the reason why until now no meaningful result has been obtained using a impure state formed out of states belonging to different domains of a self-adjoint operator. 

We now show that it is indeed possible to control the divergence using RG techniques leading to finite expectation values of the observables in the impure state.

We observe first that the expectation value of the Hamiltonian in (\ref{n-level-energy}) depends on the self-adjoint parameter $\theta$, the ultraviolet cut-off $K$ and the radius of the circle $R$. From (\ref{n-level-energy}), we observe that the fixed $n$-th level energy $E_n=\langle H \rangle_n$ diverges as $K \rightarrow \infty$ or $R \rightarrow 0$, or  both.  We next consider two of these cases separately.

{\bf $R$ fixed while $K \rightarrow \infty$}: This situation corresponds to the particle on a circle of fixed radius $R$. In this case, as the UV cut-off is removed, that is, $K \rightarrow \infty$,  the expectation value of the Hamiltonian in the impure state constructed out of the level $n=0$, given by
\begin{equation}
	\label{ground-state-1}
\langle H \rangle_0 = \lim_{K\to\infty}~  \frac{1}{2R} \left[  \frac{\theta^2}{4\pi^2} + \frac{1-\cos 2 \theta}{\pi^2}~ K \right],
\end{equation}
clearly diverges. In any physical system we must have a finite expectation value of the Hamiltonian in the ground state. In order to achieve that we now propose a RG flow in the parameter $\theta$. Therefore we promote the self-adjoint parameter $\theta$ to a function of the ultraviolet cut-off $K$. We then demand that as $K \rightarrow \infty$, the expectation value of the Hamiltonian in the ground state becomes independent of the cut-off \cite{jackiw,rajeev}. This leads to the condition
\begin{equation}
	\label{ground-energy-cond-1}
\lim_{K\to\infty} \frac{d \langle H \rangle_0}{d K} = 0.
\end{equation}
We find from (\ref{ground-state-1}) that the corresponding $\beta$-function is
\begin{equation}
  \beta_K(\theta)\equiv K\frac{d\theta}{dK}=2K~\frac{\cos 2\theta - 1}{\theta+4K\sin 2\theta}.
\end{equation}
For small $\theta$, such that $\sin2\theta/2\theta\to 1$ and $\cos2\theta\to 1-2\theta^2$, this equation can be integrated to give
\begin{equation}
	\label{theta-after-int-1}
  \theta=\frac{C}{\sqrt{1+4K}} \approx \frac{C}{\sqrt{4K}},
\end{equation}
with $C$ a constant of integration. The constant $C$ is not predicted by the theory but must be obtained empirically. For example, if the value of the self-adjoint extension parameter $\theta$ can be measured to have the value $\theta_0$ at a given scale $K = K_0$, then $C = {\sqrt{4K_0}} \theta_0$.

The above RG flow, for small values of $\theta$, renders the expectation value of the Hamiltonian in the impure state comprising of the $n=0$ level finite, as the UV cut-off $K$ is taken to infinity while $R$ is fixed. Indeed, substituting (\ref{theta-after-int-1}) into (\ref{ground-energy-cond-1}), then taking the limit $K\to\infty$, we obtain
\begin{equation}
  \langle H \rangle_0=\frac{C^2}{8\pi^2 R}.
\end{equation} 
Under this RG flow, the ratio of the expectation values of the Hamiltonian in the excited state to that in the ground state also remains finite, that is, 
\begin{equation}
\frac{\langle H \rangle_n}{\langle H \rangle_0} = 1 + \frac{4 \pi^2 n^2}{C^2}.
\end{equation}
In our approach, the expectation value of the Hamiltonian in the ground state is traded in favour of the RG flow given by the $\beta$-function and the expectation values in the excited states are measured in terms of that in the ground state energy.

{\bf $R \rightarrow 0$ and $K \rightarrow \infty$}: Let us now consider the double limit $K \rightarrow \infty$ and $R \rightarrow 0$ at the same time. The first limit, namely $K \rightarrow \infty$ is natural, since it was introduced as a regulator and the physical quantities should finally be independent of such regulators. The limit $R \rightarrow \infty$ stands in a different footing. Indeed, the problem of a particle on a circle does not require this limit and for that problem, we are free to choose any fixed value of $R$. 

We can nevertheless study a different physical problem. For example, consider an infinitesimally thin topological defect such as a flux tube, carrying magnetic flux passing perpendicular to a plane. In order to study dynamics of particles at the edge of this defect, we first regularize the problem, by considering a flux tube of finite radius $R$. We study the dynamics in the region beyond the radius $R$ and finally take the limit $R \rightarrow 0$. For this problem, the radius $R$ serves now also as a regulator and as such we need to remove it at the end of the analysis. Observe however that in real system no defect is infinitely thin but has a finite radius depending on the scale at which it is being probed. We now see that a suitable RG flow with respect to the quantity $R$ allows us to study this situation.

In this case, the expectation value of the Hamiltonian in the impure state constructed from $n=0$ level states, which is given by (\ref{ground-state-1}), now diverges as both $R \rightarrow 0$ and $K \rightarrow \infty$. In the same spirit as before, we can now consider the self-adjoint extension parameter $\theta$ as a function of both $K$ and $R$, that is, $\theta \equiv \theta (K,R)$ and demand that
\begin{equation}
 \lim_{K\to\infty} \frac{\partial \langle H \rangle_0}{\partial K} = 0, \qquad \lim_{R\to 0} \frac{\partial \langle H \rangle_0}{\partial R} = 0.
\end{equation}

We again assume that $\theta$ is small and also that it can be factorized as $\theta (K, R) = \theta_1 (K) ~ \theta_2 (R)$. This gives the corresponding $\beta$-functions
\begin{align}
\beta_K (\theta_1) &= K \frac{\partial \theta_1 (K)}{\partial K} =   - \frac{\theta_1}{2}, \\
\beta_R (\theta_2) &= -R \frac{ \partial \theta_2(R)}{\partial R} =   - \frac{\theta_2}{2},
\end{align}
which can be integrated to
\begin{equation}
\theta_1(K) = \frac{C_1}{\sqrt{4K}}, \qquad \theta_2(R) = C_2 {\sqrt{R}}.
\end{equation}
As before the constants $C_1$ and $C_2$ have to be determined empirically. If at a given $K=K_0$ and $R=R_0$, we have $\theta_1 = \theta_1(K_0),~~\theta_2 = \theta_2(R_0)$, we find that $C_1 = \theta_1(K_0) {\sqrt{4K_0}}$ and $C_2 = \theta_2(R_0)/\sqrt{R_0}$.
 
These flows of the self-adjoint extension parameter, in the limit of $K \rightarrow \infty$ and $R \rightarrow 0$, render $\langle H \rangle_0$ finite, that is, 
\begin{equation}
 \langle H \rangle_0 = \frac{C_1^2C_2^2}{4 \pi^2}.
\end{equation}
The ratio of the expectation values of the Hamiltonian in the excited state to that in the ground state is given by
\begin{equation}
\frac{\langle H \rangle_n}{\langle H \rangle_0} = 1 + \frac{4 \pi^2}{C_1^2 C_2^2}\frac{n^2}{R}. 
\end{equation}

We observe that at any value $R$, the system is gapped and the gap increases as $R \rightarrow 0$. This may be physically interpreted as follows. Suppose there exists a topological defect such as a vertex operator or a flux tube passing perpendicular to a plane. Suppose further that we aim at analyse the dynamics of parity $P$ and time-reversal $T$ invariant excitations at the boundary of these defects. Insistence on these symmetries leads us to the use of impure states. Naively the energies associated to such impure states would be infinite as the UV cut-off $K$ is removed. Our renormalization procedure gives a completely well-defined answer to this problem when $K \rightarrow \infty$. Now, in addition we may analyse also the nature of the such excitations as the radius of the topological defect changes, that is, it is governed by the RG flow associated to the radius $R$. For any radius $R$, these ``edge'' excitations are gapped and the gap increases as $R \rightarrow 0$.


The need for impure states to restore anomalous symmetries may be seen to arise from a different perspective \cite{Balachandran2013}, which is intimately related to the emergence of entanglement or von Neumann entropy due to partial observations.

The quantum observables of the particle on a circle form an algebra $\mathcal{A}$ generated by operators $\exp\left(i p_\varphi \hat{\varphi}\right)$ and $\exp\left(i \varphi \hat{p}_\varphi \right)$. A state $\omega$ on this algebra is a non-negative linear functional compatible with adjoint conjugation and normalizable to $1$. For any observable $\mathcal{O}\in\mathcal{A}$, one may associate the expectation value $\omega(\mathcal{O})\equiv \langle \mathcal{O} \rangle_{\rho}=\Tr~\rho \mathcal{O}$, where $\rho$ is some density matrix. 
  
Under time-reversal $T$ (or parity $P$), the algebra $\mathcal{A}$ may be written as $\mathcal{A}_+\oplus \mathcal{A}_-$, where $\mathcal{A}_{\pm} =\{ a\in\mathcal{A}, ~ T a T = \pm a\}$. 
The subalgebra of $T$-even observables is $\mathcal{A}_{\rm{even}}=\mathcal{A}_+\oplus \mathbb{C}\mathbb{P}_-$, where $\mathbb{P}_-$ is an orthonormal  projector ($\mathbb{P}_-^2=\mathbb{P}_-$) to $\mathcal{A}_-$ such that $\mathbb{1}=\mathbb{P}_++\mathbb{P}_-$and $\mathbb{C}$ denotes complex numbers.


Consider a state $\omega_\theta$ such that under $T$ it becomes $\omega_{-\theta}$, that is, the associated density matrix $\rho_\theta$ satisfies $T\rho_\theta T=\rho_{-\theta}$. Then the restriction of such state to the subalgebra of $T$-even observables $\mathcal{A}_{\rm{even}}$ is equivalent to considering the impure state $\omega=\omega_\theta+\omega_{-\theta}$ on the algebra $\mathcal{A}$. Indeed, if $\mathcal{O}_{\rm{odd}}\notin \mathcal{A}_{\rm{even}}$, then $\omega(\mathcal{O})=\left(\omega_\theta+\omega_{-\theta}\right)\left(\mathcal{O}_{\rm{odd}} \right)=0$.


The Hamiltonian (\ref{hamiltonian-circle}) is only formally $T$- and $P$-invariant. As we have seen before, a quasi-periodic boundary condition defining the domain of $H$ breaks these symmetries. Therefore the self-adjoint Hamiltonian $H$ is neither $T$- or $P$-even nor $T$- or $P$-odd. Nevertheless, $H$ being an observable in $\mathcal{A}$ may be decomposed as an $T$-even and $T$-odd components, that is, $H=H_{\rm{even}} + H_{\rm{odd}}$.  To see this, note that
\begin{equation}
  H_\theta=\frac{1}{R}\left(p_\varphi+\frac{\theta}{2\pi} \right)^2=\frac{1}{R}\left(p_\varphi^2+\frac{\theta^2}{4\pi^2} + \frac{\theta}{\pi}p_\varphi \right).
\end{equation}
The even and odd components of this Hamiltonian are
\begin{align}
  H_{\rm{even}} =\frac{1}{R}\left( p_\varphi^2+\frac{\theta^2}{4\pi^2} \right), \qquad
  H_{\rm{odd}} = \frac{1}{R}\frac{\theta}{\pi} p_\varphi.
\end{align}
Under $T$-even partial observations, only the first component $H_{\rm{even}}$ gives a non-zero contribution. Thus, the expectation value of this Hamiltonian under a $T$-even state gives 
\begin{align}
  \omega\left(\mathbb{P}_{\rm{even}}~H~\mathbb{P}_{\rm{even}}\right) = \frac{1}{R}\left(\omega\left(p_\varphi^2 \right)  + \frac{\theta^2}{4\pi^2}\omega\left(\mathbb{P}_{\rm{even}}\right)\right).
\end{align}
 Therefore, from $\rho$ constructed out of a $n$-level eigenstate of $p_\varphi^2$ leads to 
 \begin{equation}
   \langle H_{\rm{even}} \rangle_n = \frac{1}{R}\left( n^2+ \frac{\theta^2}{4\pi^2} \langle \mathbb{P}_{\rm{even}} \rangle_n \right).
 \end{equation}
This equation is equivalent to equation (\ref{n-level-energy}) a part from a $\theta/2\pi$-translation  in $n$, that is, $n\mapsto n+\theta/2\pi$.

The mechanism of anomaly restoration discussed in this work may have applications to time-reversal and/or parity invariant edge states. Edge states appear in condensed matter systems with boundaries where the bulk is gapped. There is an intimate relationship between  anomalies and edge states. The quantum Hall effect \cite{ezawa} is a well known example of this relationship, where the $T$ symmetry is broken. A natural question is whether similar anomaly/edge states relationship can be realized in systems with  $T$ and/or  $P$ invariant edge states. For instance, quantum spin Hall (QSH) samples \cite{kane2005z_} are in the universality class of systems with gapped bulk and $T$ invariant edge states. The quantum states of a particle on a circle may serve to model certain classes of edge states in such systems. In particular, the partial observation of $T$-even observables would induce the emergence of $T$ even impure states in the edge. The associated entanglement entropy would provide evidence for the mixed state. Such partial observations may be a result of externally choices of what one is attempting to measure in such samples. Nevertheless, it may also be argued that some couplings among the microscopic degrees of freedom may naturally reproduce such partial observations. Examples of such couplings seem to be the spin-orbit couplings \cite{kane2005z_} or coupling leading to instanton-like terms in the effective Hamiltonian \cite{Balachandran2012h}.

\section*{Acknowledgement}

KSG and ARQ would like to thank Prof. Alvaro Ferraz (IIP-UFRN-Brazil)
for the hospitality at IIP-Natal-Brazil, where this work was carried
out. We also thank A. P. Balachandran for discussions. 
ARQ is supported by CNPq under process number 305338/2012-9.


\begin{thebibliography}{99}

\bibitem{Harvey2007}
J.~A. Harvey, {\it Tasi 2003 lectures on anomalies},
  \href{http://xxx.lanl.gov/abs/hep-th/0509097}{{\tt hep-th/0509097}}.

\bibitem{jackiw1969} J. S. Bell and R, Jackiw, Nuovo Cim. {\bf A60}, 47 (1969).

\bibitem{Adler1969a}
S.~L. Adler,   {\em Phys.Rev.} {\bf 177}, (1969) 2426.

\bibitem{Fujikawa1984d}
K.~Fujikawa,  {\em Phys.Rev.} {\bf D29}, (1984) 285.

\bibitem{Adler1969}
S.~L. Adler and W.~A. Bardeen,  {\em Phys.Rev.} {\bf 182}, 1517 (1969).

\bibitem{Alvarez-Gaume1984}
L.~Alvarez-Gaume and P.~H. Ginsparg, {\em Nucl.Phys.} {\bf B243}, 449 (1984).

\bibitem{AlvarezGaume1984269}
L.~Alvarez-Gaume and E.~Witten,   {\em Nucl. Phys. } {\bf B234}, 269 (1984).

\bibitem{Witten1982324}
E.~Witten,   {\em Phys. Lett. } {\bf B117}, 324 (1982).

\bibitem{Bimonte1993}
G.~Bimonte, Kumar S. Gupta, and A.~Stern,   {\em Int. J. Mod. Phys.} {\bf A8}, 653 (1993).
  

\bibitem{Balachandran1996a}
A.~P. Balachandran, L.~Chandar, and A.~Momen,  {\em Nucl. Phys.} {\bf B461}, 581 (1996).  

\bibitem{CallanJr1985427} C. Callan and J. Harvey,  Nucl. Phys. {\bf B250}, 427 (1985).

\bibitem{Stone198789} M. Stone and F. Gaitan,  Ann.  Phys. {\bf 178}, 89 (1987).

\bibitem{Balachandran1992}
A.~P. Balachandran, G.~Bimonte, Kumar~S. Gupta, and A.~Stern,  {\em Int. J. Mod. Phys.} {\bf A7}, 4655 (1992).

\bibitem{Balachandran1995b}
A.~P. Balachandran, L.~Chandar, and E.~Ercolessi,  {\em Int. J. Mod. Phys.} {\bf A10}, 1969 (1995).

\bibitem{Balachandran1994a}
A.~P. Balachandran, L.~Chandar, E.~Ercolessi, T.~R. Govindaraj~an, and R.~Shankar,   {\em Int. J. Mod. Phys.} {\bf A9}, 3417 (1994). 

\bibitem{reed1975methods}
M.~Reed and B.~Simon, {\em Methods of modern mathematical physics: Fourier
  analysis, self- adjointness}, vol.~2. \newblock Academic press, 1975.

\bibitem{PhysRevD.34.674}
J.~G. Esteve,  {\em Phys. Rev. } {\bf D34}, 674 (1986).

\bibitem{jackiw} R. Jackiw, M. A. B. Beg Memorial Volume, A. Ali and P. Hoodbhoy Eds. (World Scientific, Singapore, 1991).

\bibitem{PhysRevD.66.125013}
J.~G. Esteve, {\em Phys. Rev.} {\bf D66}, 125013 (2002).

\bibitem{PhysRevLett.87.220402}
H.~E. Camblong, L.~N. Epele, H.~Fanchiotti, and C.~A. Garc\'\i{}a~Canal,  {\em Phys. Rev. Lett.} {\bf 87}, 220402 (2001).

\bibitem{ksgpolar} P. R. Giri, Kumar S. Gupta, S. Meljanac and A. Samsarov, Phys. Lett. {\bf A372}, 2967 (2008). 

\bibitem{ezawa} {\it Quantum Hall Effects}, Z. F. Ezawa, World Scientific (2008).

\bibitem{biru1} B. Basu-Mallick, Pijush K. Ghosh and Kumar S. Gupta, Nucl. Phys. {\bf B659}, 437 (2003).

\bibitem{biru2} B. Basu-Mallick, Pijush K. Ghosh and Kumar S. Gupta, Phys. Lett. {\bf A311}, 87 (2003).

\bibitem{biru3} B. Basu-Mallick, Kumar S. Gupta, S. Meljanac and A. Samsarov, Eur. Phys. Jour. {\bf C53}, 295 (2008).

\bibitem{sen1} D. Birmingham, Kumar S. Gupta and S. Sen, Phys. Lett. {\bf B505}, 191 (2001).

\bibitem{sen2} Kumar S. Gupta and S. Sen, Phys. Lett. {\bf B526}, 121 (2002).

\bibitem{Balachandran2012}
A.~P. Balachandran and A.~R. Queiroz, {\em Phys. Rev.} {\bf D85} 025017 (2012).
  
\bibitem{Balachandran:2011gj}  A.~P. Balachandran and A. R. de Queiroz,  {\em JHEP} {\bf 11}, 126 (2011). 

\bibitem{rajeev} Kumar S. Gupta and S. G. Rajeev, Phys. Rev. {\bf D48}, 5940 (1993).

\bibitem{Balachandran2013}
A.~Balachandran, T.~Govindarajan, A.~R. de~Queiroz, and A.~Reyes-Lega, Phys. Rev. Lett. {\bf 110}, 080503 (2013).
  

\bibitem{kane2005z_}
C.~Kane and E.~Mele,  {\em Phys. Rev. Lett} {\bf 95}, 146802 (2005).

 \bibitem{Balachandran2012h} A. P. Balachandran, T. Govindarajan, and A. R. de Queiroz,  Eur.Phys.J.Plus {\bf 127}, 118 (2012).

\end{thebibliography}
\end{document}